\documentclass[conference]{IEEEtran}
\IEEEoverridecommandlockouts
\usepackage{cite}
\usepackage{amsmath,amssymb,amsfonts}
\usepackage{algorithmic}
\usepackage{graphicx}
\usepackage{textcomp}
\usepackage{xcolor}
\def\BibTeX{{\rm B\kern-.05em{\sc i\kern-.025em b}\kern-.08em
    T\kern-.1667em\lower.7ex\hbox{E}\kern-.125emX}}

\IEEEoverridecommandlockouts
\usepackage{bm}
\usepackage{cite}
\usepackage{amsmath,amssymb,amsfonts}
\usepackage{algorithmic}
\usepackage{graphicx}
\usepackage{threeparttable}
\usepackage{textcomp}
\usepackage{algorithmic}
\usepackage{algorithm}
\usepackage{url}
\usepackage{xcolor} % For coloring text
\usepackage{times}  % for Times font
\usepackage{mathptmx}

\makeatother

\def\BibTeX{{\rm B\kern-.05em{\sc i\kern-.025em b}\kern-.08em
    T\kern-.1667em\lower.7ex\hbox{E}\kern-.125emX}}
\begin{document}

\title{Robust Model-Free Control Framework with Safety Constraints for a Fully Electric Linear Actuator System\\

\thanks{Funding for this research was provided by the Business Finland partnership project "Future All-Electric Rough Terrain Autonomous Mobile Manipulators" (Grant No. 2334/31/2022).}% <-this % stops a space
}

\author{\IEEEauthorblockN{1\textsuperscript{st} Mehdi Heydari Shahna}
\IEEEauthorblockA{\textit{Engineering and Natural Sciences} \\
\textit{Tampere University}\\
Tampere, Finland \\
mehdi.heydarishahna@tuni.fi}
\and
\IEEEauthorblockN{2\textsuperscript{nd} Pauli Mustalahti}
\IEEEauthorblockA{\textit{Engineering and Natural Sciences} \\
\textit{Tampere University}\\
Tampere, Finland \\
pauli.mustalahti@tuni.fi}
\and
\IEEEauthorblockN{3\textsuperscript{rd} Jouni Mattila}
\IEEEauthorblockA{\textit{Engineering and Natural Sciences} \\
\textit{Tampere University}\\
Tampere, Finland \\
jouni.mattila@tuni.fi}

}

\maketitle

\begin{abstract}
This paper introduces a novel model-free control strategy for a complex multi-stage gearbox electromechanical linear actuator (EMLA) system, driven by a permanent magnet synchronous motor (PMSM) with non-ideal ball screw characteristics. The proposed control approach aims to (1) manage user-specified safety constraints, (2) identify optimal control parameters for minimizing tracking errors, (3) ensure robustness, and (4) guarantee uniformly exponential stability. First, this paper employs a trajectory-setting interpolation-based algorithm to specify the piecewise definition of a smooth and jerk-bounded reference trajectory. Then, a dual robust subsystem-based barrier Lyapunov function (DRS-BLF) control is proposed for the PMSM-powered EMLA system to track the reference motions,
guaranteeing user-specified safety related to constraints on system characteristics and alleviating control signal efforts. This methodology guarantees robustness and uniform exponential convergence. Lastly, optimal control parameter values are determined by customizing a swarm intelligence technique known as the Jaya (a term derived from the Sanskrit word for `victory') algorithm to minimize tracking errors. Experimental results validate the performance of the DRS-BLF control.
\end{abstract}

\vspace{0.2cm}

\section{Introduction}
\subsection{Background of Fully Electric Actuation Systems}
Global agreements, such as the 2015 Paris Agreement \cite{agreement2015paris}, underscore the immediate need for a full transition to electricity, as emphasized in \cite{moghaddasi2021net}. This transition has also influenced the heavy machinery industry to develop electrified equipment, aiming to enhance environmental performance. Historically, heavy-duty manipulators have depended on hydraulic actuators (HAs) to transfer energy and produce movement. While HAs have strengths, they face challenges, including high maintenance, environmental risks, and reduced performance \cite{qu2023electrified}. Embracing sustainable practices, Bobcat has produced the all-electric compact loader T7X \cite{bobcat}, while Volvo has launched the Volvo L25 Electric \cite{volvo}, marking the start of a new series of electric compact wheel loaders. In addition, Ponsse and Epec have unveiled the EV1, an innovative electric forest machine technology concept in the forestry sector \cite{ponsse}. Further, Sandvik has introduced the Toro™ LH625iE, recognized as the world's largest underground loader \cite{sandvik}.
Thus, electromechanical actuators (EMAs) are revolutionizing future actuation systems in the heavy-duty machinery industry for such tasks as lifting and moving heavy objects \cite{badrinarayanan2018electro, crowder2019electric}. These actuators can be designed in either linear or rotary forms \cite{todeschi2015health}, where screw-based electromechanical linear actuators (EMLAs) allow the efficient conversion of rotary motion into linear displacement, essential for tasks requiring high levels of push or pull force.
EMLAs offer distinct advantages in terms of mechanical simplicity, robustness, efficiency, and cost-effectiveness, rendering them particularly suitable for heavy-duty machinery applications \cite{fleming2021electrification}. In addition, their streamlined design, featuring fewer moving components, results in diminished maintenance demands compared to electro-hydraulic actuators (EHAs), thus addressing such issues as oil leakage, commonly associated with conventional hydraulic systems.
Meanwhile, permanent magnet synchronous motors (PMSMs) are particularly favored for integration within EMLAs due to their exceptional efficiency and notable torque density \cite{cavallaro2005efficiency}.
\begin{figure}[h!]
    \hspace*{-0.0cm} % Adjust the value as needed
    \centering
    \scalebox{1}{\includegraphics[trim={0cm 0.0cm 0.0cm 0cm},clip,width=\columnwidth]{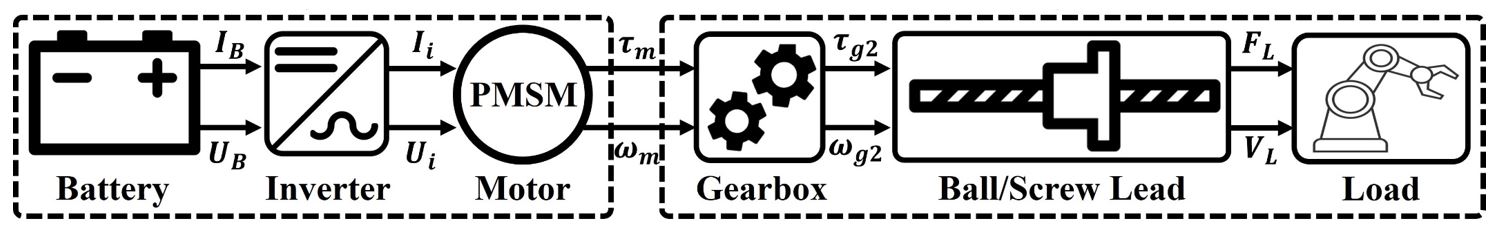}}
    \caption{Configuration of the EMLA system driven by a PMSM \cite{heydari2024robust}.}
    \label{fig1111}
\end{figure}
\\
Fig. \ref{fig1111} depicts the configuration of an EMLA system powered by a PMSM, which converts electrical voltage and current into angular torque and velocity and, subsequently, into linear force and velocity \cite{heydari2024robust}. The motor functions as the primary source of rotational power, while the lead, ball, or roller screw transforms rotational motion into linear movement. 

\subsection{Control Challenges and Motivations of the Paper}
This technology transition is not without its challenges for PMSM-powered EMLAs, which can be itemized as follows:
\subsubsection{Non-ideal configurations and modeling errors}
Although engineers can attain accurate multi-stage EMLA parameter values through careful calibration, system identification techniques, and other strategies, such factors as manufacturing variations, temperature, operating conditions, measurement equipment limitations, and motor model complexity can degrade control performance \cite{li2022design}. More precisely, we delineate some technical reasons for modeling errors, as follows: 
\begin{itemize}
    \item Inductance configurations of PMSMs: The physical manufacturing imperfections in motor construction, such as variations in magnet placement or geometric irregularities, can lead to differences in inductances along the d-q axis ($L_q$ and $L_d$) \cite{desai2023novel}, known as `asymmetrical motor.' 
    \item Characteristics of the ball screw: A ball screw may exhibit nonlinear and non-ideal characteristics, impacting dynamic behavior \cite{caracciolo2014optimal}.
    \item Mechanical stiffness: This directly influences the physical dynamics of a system, including how precisely the system can be controlled, its stability, and its response to external forces, particularly in tasks where deformation under load affects functionality \cite{vesterinen2022conseptual}.
    \item Numerical stiffness: This is primarily relevant in the context of simulations used for designing and testing control systems, and it includes the need for finer discretization in simulations, which can slow computations and impact the feasibility of real-time control implementations.
    \item Latent uncertainty and disturbance: Flux harmonics within the PMSM due to a non-uniform flux density distribution in the air gap and fluctuations or abnormalities in the electrical supply and the switching operation of the inverter can cause external disturbances \cite{yang2016disturbance,de2017robust}.
\end{itemize}

\subsubsection{User-specified safety constraints}
The voltage supplied to the motor's terminals dictates the stator current, which in turn adjusts the torque generated by the PMSM to meet the desired level \cite{preindl2014optimal}. Considering constraints on all these signals, as well as on the motion states of the EMLA, is crucial to protecting the motor and associated electronics, including inverters, from potential harm, maintaining compliance with industry standards, and protecting overall system integrity and human safety \cite{yang2016disturbance, heydari2024robust}.

\subsubsection{Tuning control parameters}
traditionally, the process of achieving effective control gains in actuation systems relies on computationally intensive solutions \cite{shahna2024obstacle}, such as:
\begin{itemize}
\item Trial and error: This method is inherently time-consuming and typically conducted in an ad-hoc manner. It involves repeatedly adjusting control parameters based on observed performances until satisfactory results are achieved. However, this approach lacks a systematic methodology and can lead to a suboptimal performance if not meticulously managed \cite{moudgal1995fuzzy}.
\item General guidelines or experience: This approach involves adopting recommended parameter settings from the existing literature or utilizing insights gained from similar applications. While this can provide a starting point, it often fails to deliver optimal results in unique situations where specific experience is lacking or when the systems involved have distinct characteristics or operate under conditions previously unencountered \cite{wang2012machine}.
\end{itemize}
\subsection{Literature review}
In references \cite{xia2023prescribed} and \cite{song2023barrier}, the authors introduce a novel control method to handle state constraints using Barrier Lyapunov Functions (BLF) for PMSMs. A Barrier Lyapunov candidate is a function that approaches infinity as its argument nears a predefined safety constraint limit.
Although the BLF-based controller is a mathematically elegant way to handle state constraints systematically, it often demands significant control efforts when the state trajectory approaches the safety boundary \cite{garg2020barrier}. In addition, heuristic algorithms are popular methods for optimizing control gains because they quickly find reasonably good solutions for complex problems, making them practical for various real-world applications \cite{rodriguez2021self, jin2021self}. The comprehensive reviews in \cite{houssein2021JAYA} and \cite{bansal2021single} demonstrate that the Jaya algorithm consistently outperformed other established swarm-based algorithms, exhibiting significantly better results in terms of both accuracy and convergence speed.
Much like teaching learning-based optimization (TLBO), proposed in \cite{rao2011teaching}, the Jaya algorithm does not require specific algorithmic parameters. Instead, it operates with only common parameters, such as population size and a maximum number of generations \cite{rao2019JAYA}. Furthermore, it eliminates the need for a learner phase, utilizing only a single phase, the teacher phase, whereas TLBO executes its operations in two distinct phases.
\subsection{Paper Contributions and Organization}
In this paper, we propose a dual-robust subsystem-based barrier Lyapunov function (DRS-BLF) control framework
for a PMSM-powered EMLA mechanism to cover the following control aspects of the drive system:
\begin{itemize}
    \item Current control: This would involve managing the current within the motor coils to ensure sufficient torque generation and performance, serving as the base layer of the motor control hierarchy.
    \item Motion control: We also focus on motion control, particularly how the actuator manages and regulates the movement of the linear drive in tracking both the smooth desired position and velocity simultaneously.
\end{itemize}
By using this comprehensive approach, the proposed control aims to adhere to safety constraints specified by the user in terms of EMLA motion, including linear position and velocity, as well as defined motor characteristics, including current, torque, and voltage. This paper has several contributions to drive system control:
\begin{itemize}
    \item The DRS-BLF control framework is proposed to address not only EMLA motion control but also the current control of the PMSM to track defined control tasks.
    \item By using this approach, the proposed control adheres to safety constraints specified by the user in terms of EMLA motion and PMSM characteristics.
    \item A mathematical saturation function is incorporated to reduce torque and voltage efforts, which is a widespread problem in the BLF control concept \cite{garg2020barrier}.
    \item DRS-BLF parameters are optimized using a high-performance swarm intelligence technique, the Jaya algorithm, to minimize tracking errors.
    \item This control framework guarantees robustness and uniform exponential stability of the EMLA framework.
\end{itemize}
\indent After this section, we delve into the necessary mathematical details of PMSM-driven EMLA modeling and constraints on control efforts in Section II. Section III is dedicated to describing jerk-bounded and smooth trajectory planning based on quintic polynomials. Section IV provides a step-wise design of the DRS-BLF control for the EMLA mechanism. Section V demonstrates that applying the DRS-BLF control to the EMLA guarantees uniformly exponential stability under modeling errors and disturbances. In Section VI, we illustrate the use of Jaya optimization for tuning DRS-BLF control gains to minimize tracking errors. Finally, we investigate the effectiveness of the DRS-BLF control through experiments on an EMLA system driven by a PMSM in section VII.
\section{Pmsm-powered Emla Modeling }
\subsection{Mathematical EMLA Model} 
This section presents an EMLA mechanism driven by a surface-mounted PMSM to convert electric power to mechanical power. To eliminate the dependence of the PMSM's three-phase voltage on the rotor angle, the Park transformation can be utilized \cite{heydari2024robust}. Subsequently, the EMLA torque signal produced by the PMSMs can be expressed as:
\begin{equation}
\small
\tau_{m_{}} = \frac {3}{2} P_{} \left( i_{q_{}} \left( \theta_{{PM}_{}} + (L_{d_{}} - L_{q_{}}) i_{d_{}} \right) \right)
\label{equation:1}
\end{equation}
where $\theta_{PM}$ is the flux linkage, $P_{}$ is the number of rotor pole pairs, and $L_{d}$ and $L_{q}$ are unknown inductances in the $d$-axis and $q$-axis. The two currents of the PMSM in the $d$-axis and $q$-axis are represented as $i_{d}$ and $i_{q}$. 
We consider sufficient torque $\tau^{*}_{m_{}}$ and currents $i^*_{q}$ and $i^*_{d}$ as references for the EMLA to perform a motion task. By setting $i^*_{d}=0$, the magnetic field along the $d$-axis is eliminated, and the torque generated by the electric machine becomes independent of the rotor's position. As a result, the load position can be controlled without affecting the magnetic field along the $d$-axis. The $q$-axis current, $i_{q}$, can then be used to generate the torque necessary for motion control. Thus, the reference electromagnetic torque can be simplified, as in \cite{heydari2024robust}:
\begin{equation}
\small
\begin{aligned}
 \tau^*_{m} = \frac {3}{2} {P}{i^*_{q}}\theta_{{PM}}
\label{equation:4}
\end{aligned}
\end{equation}
The electromagnetic torque of each electric motor can be written as the function of linear motion characteristics at the load side of the EMLA (${\ddot{x}_{L}\in \mathbb{R}}$, ${\dot{x}_L \in \mathbb{R}}$, and ${{x}_L \in \mathbb{R}}$) \cite{heydari2024robust}:
\begin{equation}
\small
\begin{aligned}
 \tau^*_{m} = I_{{eq}}\ddot{x}_{L}+{B_{{eq}}}\dot{x}_{L}+\epsilon_{{eq}}x_{L}+f_{{eq}}F_{L}
\label{equation:2}
\end{aligned}
\end{equation}
where $I_{{eq}}$, $\epsilon_{{eq}}$, $B_{{eq}}$, and $f_{{eq}}$ are the equivalent inertia, stiffness, equivalent viscosity, and the force coefficient of the entire EMLA system, respectively. As well, $F_{L}$ is the load force, and $x_{L}$, $\dot{x}_{L}$, and $\ddot{x}_{L}$ are the linear position, velocity, and acceleration of the screw mechanism at the load side, respectively. In terms of the EMLA's linear motion at the load side, the equations for the derivatives of the motor currents can be obtained as follows:
\begin{equation}
\small
\begin{aligned}
 &\dot{i}_{q} = {\frac{1}{L_{q}}}{(u_{q}}-{R_{s}}{i_{q}}-{P}{{c}_{{RL}}}{\dot{x}_{L}}^{\top}{L_{d}}{i_{d}}-{P}{{c}_{{RL}}}{\dot{x}_{L}}{\theta_{{PM}}}) \\
 &\dot{i}_{d} = {\frac{1}{L_{d}}}{(u_{d}}-{R_{s}}{i_{d}}+2{P}{{c}_{{RL}}}{\dot{x}_{L}}{L_{q}}{i_{q}})
\label{equation:3}
\end{aligned}
\end{equation}
where $c_{{RL}}$ is the coefficient of converting rotary movement to linear movement and ${R_{s}}$ is the resistance of the stator winding. In addition, the two voltages of the PMSM along the $d$-axis and $q$-axis have been donated by $u_{q}$ and $u_{d}$. In this study, we do not assume the rotor of the surface-mounted PMSM has a completely ideal non-salient shape. Thus, the inductances $L_{d}$ and $L_{q}$ are non-identical. By transferring \eqref{equation:2} and \eqref{equation:3} into a subsystem-based form, we have:
\begin{equation}
\small
\begin{aligned}
\label{equation:5}
&\dot{x}_{1}=x_{2}\\
&\dot{x}_{2} = \frac{1}{I_{{eq}}} u_{2}+\frac{B_{{eq}}}{I_{{eq}}}x_{2}+\frac{\epsilon_{{eq}}}{I_{{eq}}}x_{1}+\frac{f_{{eq}}F_{L}}{I_{{eq}}}\\
&\dot{x}_{3} = \frac{1}{L_{q}}{u_{3}}-\frac{R_{s}}{L_{q}}{x_{3}}-\frac{{P}{{c}_{{RL}}}{L_{d}}}{L_{q}}{x_{2}}{x_{4}}-\frac{{P}{{c}_{{RL}}}{\theta_{{PM}_i}}}{L_{q}}{x_{2}} \\
&\dot{x}_{4} = {\frac{1}{L_{d}}}{u_{4}}-\frac{R_{s}}{L_{d}}{x_{4}}+\frac{2{P}{{c}_{{RL}}}{L_{q}}}{L_{d}}{x_{2}}{x_{3}} \\
\end{aligned}
\end{equation}
where $u_{2}=\tau^*_{m}$, $u_{3}=u_{q}$, $u_{4}=u_{d}$, $x_{1}=x_{L}$, $x_{2}=\dot{x}_{L}$, $x_{3}=i_{q}$, and $x_{4}=i_{d}$. To simplify \eqref{equation:5}, we can express the system as:
\begin{equation}
\small
\begin{aligned}
\label{equation:6}
&\dot{x}_{1_{}}=h_{1_{}} x_{2_{}}+\Delta_{1_{}} (x_{1_{}},x_{2_{}},x_{3_{}},x_{4_{}})+d_{1_{}}\\
&\dot{x}_{2_{}} = h_{2_{}} u_{2_{}}+\Delta_{2_{}} (x_{1_{}},x_{2_{}},x_{3_i},x_{4_{}})+d_{2_{}}\\
&\dot{x}_{3_{}} = h_{3_{}} {u_{3_{}}}+\Delta_{3_{}} (x_{1_{}},x_{2_{}},x_{3_{}},x_{4_{}})+d_{3_{}} \\
&\dot{x}_{4_{}} = h_{4_{}}{u_{4_{}}}+\Delta_{4_{}} (x_{1_{}},x_{2_{}},x_{3_{}},x_{4_{}})+d_{4_{}}
\end{aligned}
\end{equation}
where \eqref{equation:5} is a system including four subsystems (four states $X=[x_{1},...,x_{4}]^{\top}$) and three control inputs. For $j=1,...,4$, we know $h_{j}$ is positive and assumed an unknown coefficient, $\Delta_{j}$ could be unknown uncertainty, and $d_{j}$ is the unknown time-variant disturbance, including load effects.
\subsection{Mathematical Model of Voltage and Torque Constraints}
We define ${Sat}(.)$ to operate in compliance with the constraints imposed on the control signals:
\begin{equation}
\small
\begin{aligned}
\label{equation: 7}
 {Sat}(u_k) = \begin{cases}u^+_k, &  u_k \geq u^+_k \\
u_k &  u^-_k \leq u_k \leq u^+_k \\
u^-_k &  u_k \leq u^-_k\end{cases} , \hspace{0.3cm} k=2,3,4
\end{aligned}
\end{equation}
where $u^+_k$ and $u^-_k$ denote the upper and lower nominal bounds, respectively, of the permissible $u_k$ values that can be generated. To elaborate further, we transform \eqref{equation: 7} as follows:
\begin{equation}
\small
\begin{aligned}
\label{equation: 8}
{Sat}(u_k)=S_{1} \hspace{0.1cm} u_k + S_{2} 
\end{aligned}
\end{equation}
where:
\begin{equation}
\small
\begin{aligned}
\label{equation: 9}
 S_{1} = \begin{cases} \frac{1}{\mid u_k \mid +1}, & \hspace{0.2cm} u_k \geq u^+_k \hspace{0.2cm}  \text{or} \hspace{0.2cm} u_k \leq u^-_k \\
1 & \hspace{0.2cm} u^-_k\leq u_k \leq u^+_k \end{cases} 
\end{aligned}
\end{equation}

\begin{equation}
\small
\begin{aligned}
\label{equation: 10}
S_{2}= \begin{cases}u^+_k - \frac{u_k}{\mid u_k \mid + 1}, & u_k \geq u^+_k \\
0 &  u^-_k \leq u_k \leq u^+_k \\
u^-_k - \frac{u_k}{\mid u_k \mid + 1} & u_k \leq u^-_k\end{cases}
\end{aligned}
\end{equation}
We have $S_{2} < \max(|u^-_k| + 1, |u^+_k| + 1)=S_{max}$, $S_{1} \leq 1$, where $S_{min}$ is the minimum of the set $S_1$. By considering \eqref{equation:6} and \eqref{equation: 8}, we can express:
\begin{equation}
\small
\begin{aligned}
\label{equation: 11}
&\dot{x}_1=h_1 x_2+\Delta_1 +d_1\\
&\dot{x}_2 = h_2 S_1 {u_2}+\Delta_2 +d_2+h_2 S_2\\
&\dot{x}_3 = h_3 S_1 {u_3}+\Delta_3 +d_3 + h_3 S_2 \\
&\dot{x}_4 = h_4 S_1 {u_4}+\Delta_4 +d_4 + h_4 S_2 \\
\end{aligned}
\end{equation}
\indent \textbf{Assumption 1.} As we know in the EMLA system, $h_j>0$ as the coefficients of the control signals are positive and bounded. Thus, there always exist some positive constant $h^-_j$ and $h^+_j$, such that $0<h^-_j \leq h_j \leq h^+_j<\infty$.
\section{Jerk-bounded Trajectory Planning}
\begin{figure}[h!]
    \hspace*{-0.0cm} % Adjust the value as needed
    \centering
    \scalebox{1}{\includegraphics[trim={0cm 0.0cm 0.0cm 0cm},clip,width=\columnwidth]{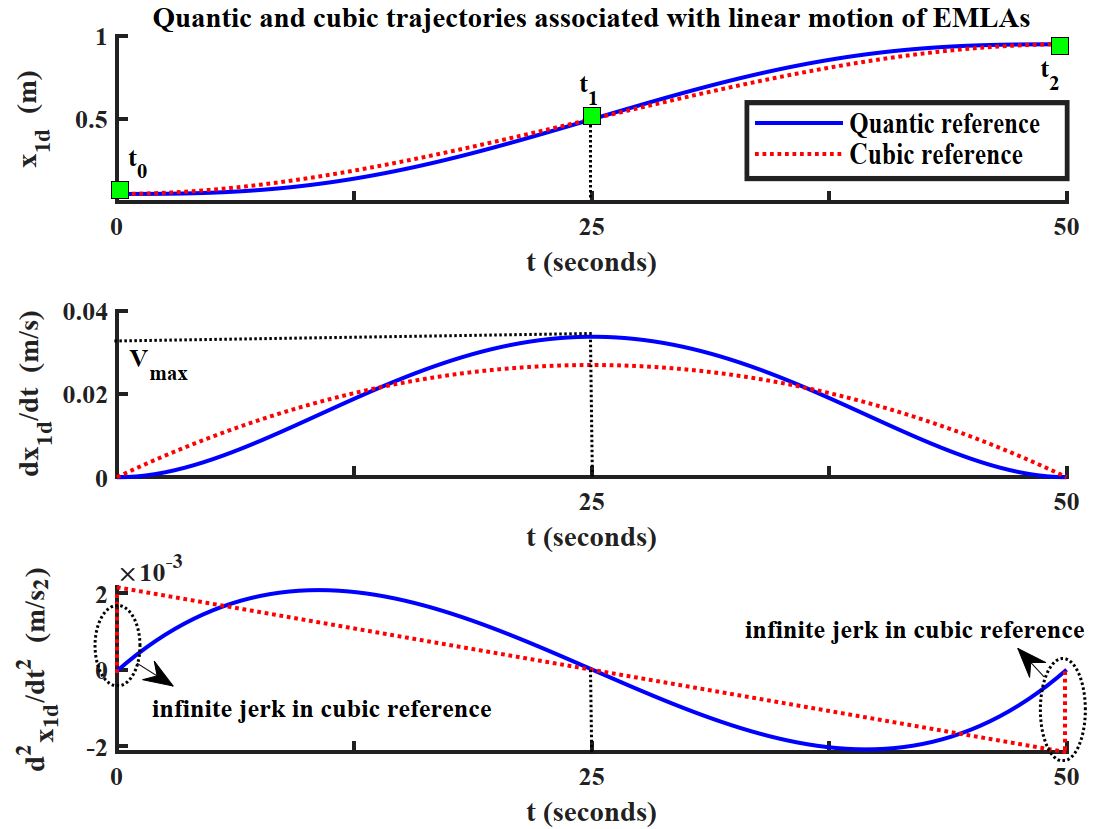}}
    \caption{Cubic and quantic polynomials under the same conditions.}
    \label{fig222}
\end{figure}

Fig. \ref{fig222} presents an example of both cubic and quantic polynomials under the same conditions. While employing cubic polynomials when interpolating trajectory settings is straightforward, it often results in an acceleration profile exhibiting discontinuities, leading to abrupt jerks. These jerks in acceleration can introduce vibrations in practical applications. In addition, the initial and final positions of the built-in trajectory based on cubic polynomials are not exactly equal to the target positions \cite{zhidchenko2023method}. To designate the start and target points for reference accelerations, a quintic polynomial is the lowest-order position reference polynomial, guaranteeing the smoothness of the jerk profile \cite{jazar2010theory}.
By designating suitable initial and target conditions for the five-order polynomials, a smooth reference trajectory between two points can be identified while maintaining continuity in acceleration. It is extendable to a series of points, each being a five-order polynomial.
Assume that $\gamma_0, \ldots, \gamma_5$ are any coefficient. In this paper, for an EMLA mechanism, we define $x_{1d}$ as a quintic polynomial as follows:
\begin{equation}
\small
\begin{aligned}
\label{equation: 14}
x_{1d}(t)=\gamma_5 t^5 + \gamma_4 t^4 + \gamma_3 t^3 + \gamma_2 t^2 + \gamma_1 t + \gamma_0
\end{aligned}
\end{equation}
By assuming $t=t_0$ and $t=t_1$ are initial and target times between two points, the following equation can be solved to find $\gamma_0,\ldots,\gamma_5$:
\begin{equation}
\small
\begin{aligned}
\label{equation: 15}
\begin{bmatrix}
1 & t_0 & t_0^2 & t_0^3 & t_0^4 & t_0^5 \\
0 & 1 & 2 t_0 & 3 t_0^2 & 4 t_0^3 & 5 t_0^4 \\
0 & 0 & 2 & 6 t_0 & 12 t_0^2 & 20 t_0^3 \\
1 & t & t_1^2 & t_1^3 & t_1^4 & t_1^5 \\
0 & 1 & 2 t_1 & 3 t_1^2 & 4 t_1^3 & 5 t_1^4 \\
0 & 0 & 2 & 6 t_1 & 12 t_1^2 & 20 t_1^3
\end{bmatrix}
\begin{bmatrix}
\gamma_0 \\
\gamma_1 \\
\gamma_2 \\
\gamma_3 \\
\gamma_4 \\
\gamma_5
\end{bmatrix}
=
\begin{bmatrix}
x_{1d} (t_0) \\
\dot{x}_{1d} (t_0) \\
\ddot{x}_{1d} (t_0) \\
x_{1d} (t_1) \\
\dot{x}_{1d} (t_1) \\
\ddot{x}_{1d} (t_1)
\end{bmatrix}
\end{aligned}
\end{equation}
By replacing $t_0$ with $t_1$, a new target point ($t_2$) can be considered to calculate the next step. Thus, a set of quintic polynomials for the whole reference trajectory can be constructed. 
\section{Drs-blf Control Framework}
The BLF is a commonly employed method for systematically devising controllers to avoid violating safety constraints during trajectory tracking. Analytical assurance of safety-constraint satisfaction is achieved by proving the boundedness of BLF in a closed loop.
Let $x_{jd}$, for $j=1,...,4$, be given desired reference states for \eqref{equation: 11}. By designing the desired trajectory based on quantic polynomials, first-time derivative $\dot{x}_{jd}$ are bounded, if we consider the tracking error, as follows:
\begin{equation}
\small
\begin{aligned}
\label{equaation: 16}
x_{ej}=& x_j - x_{jd}\\
\end{aligned}
\end{equation}
where $x_{1d}$ is the desired linear position, and $x_{2d}=\dot{x}_{1d}$ is the desired linear velocity, both provided by smooth and jerk-bounded quantic polynomials. $x_{3d}={i}^*_q$ is the desired q-axis current provided in \eqref{equation:4}, and $x_{4d}={i}^*_d=0$ is the desired d-axis current.\\
\indent \textbf{Assumption 2.}
It is assumed to satisfy:
\begin{equation}
\small
\begin{aligned}
\label{equaation: 17}
&-\chi_j < -\lambda_j \leq x_{jd} \leq \lambda_j < \chi_j, \hspace{1.3cm} j=1,3,4\\
&-\chi_j < -\lambda_j \leq x_{jd} - u_{j-1} \leq \lambda_j < \chi_j, \hspace{0.3cm} j=2\\
\end{aligned}
\end{equation}
where $\lambda_j$ and $\chi_j$ are positive constants. We introduce a transformation of tracking form by defining $e_j$, as follows:
\begin{equation}
\small
\begin{aligned}
\label{equation: 18}
 e_{j} = \begin{cases} x_{ej} & \hspace{0.2cm} j=1,3,4 \\
x_{ej}- u_j & \hspace{0.2cm} j=2 \end{cases} 
\end{aligned}
\end{equation}
where control signals $u_j$ are introduced as follows:
\begin{equation}
\small
\begin{aligned}
\label{equaation: 19}
u_j=&-\frac{1}{2}(\epsilon_j e_j+\zeta_j \hat{\theta}_j \phi_j)\\
\end{aligned}
\end{equation}
where $u_1$ is a virtual control for the linear position subsystem, and $u_2$, $u_3$, and $u_4$ have been introduced in \eqref{equation: 11}. $\epsilon_j$ and $\zeta_j$ are positive constants. $\hat{\theta}_j$ is updated by an adaptation law as:
\begin{equation}
\small
\begin{aligned}
\label{equaation: 20}
\dot{\hat{\theta}}_j=-\beta_j \kappa_j \hat{\theta}_j+\frac{1}{2} \zeta_j \beta_j \phi_j^2
\end{aligned}
\end{equation}
where $\kappa_j$ and $\beta_j$ are positive constants, and the initial condition $\hat{\theta}_j\left(t_0\right)>0$ is finite, guaranteeing $\hat{\theta}_j(t)>0$ for any $t \geq t_0$. $\phi_j$ is a positive notation and defined as:
\begin{equation}
\small
\begin{aligned}
\label{equation: 21}
\phi_j=\frac{e_j}{Q_j}, \hspace{0.2cm} Q_j=\rho_{j}^2-e_j^2
\end{aligned}
\end{equation}
where $\rho_j=\chi_j-\lambda_{j}$ is a positive constant. Assume that $\left|e_j(t)\right|<\rho_j$, as mentioned positive constants are designable.\\
\indent \textbf{Definition 1.} According to \cite{ren2010adaptive}, if $\left|e_j(t)\right|<\rho_j$, we have:
\begin{equation}
\small
\begin{aligned}
\label{equaation: 23}
\log \left(\frac{\rho_j^2}{Q_j}\right)<\frac{e_j^2}{Q_j}, \hspace{0.2cm} j=1,2,3,4
\end{aligned}
\end{equation}
\indent The function $\hat{\theta}_j$ estimates the unknown parameter $\theta_j^*$ which is defined as follows:
\begin{equation}
\small
\begin{aligned}
\label{equaation: 24}
\theta_1^*=& \frac{2 \delta_1}{\zeta_1}\left(\frac{h^+_1}{h^-_1}\right)^2+\frac{2 S_{min} \upsilon_1}{ \zeta_1}\left(\frac{\mu_1}{h^-_1}\right)^2+\frac{2 S_{min} \sigma_1}{\zeta_1}\left(\frac{{d^*_1} }{h^-_1}\right)^2\\
\theta_2^*=& \frac{Q_2^2}{2 \zeta_2 \delta_1}  +\frac{2 \upsilon_2}{ S_{min}\zeta_2}\left(\frac{\mu_2}{h^-_2}\right)^2+\frac{2 \sigma_2}{S_{min}\zeta_2}\left(\frac{{d^*_2} }{h^-_2}\right)^2\\
&+\frac{2\eta_2}{S_{min}\zeta_2}\left(\frac{S_{2} h_2}{h^-_2}\right)^2 \\
\theta_3^*=& \frac{2 \upsilon_3}{ S_{min}\zeta_3}\left(\frac{\mu_3}{h^-_3}\right)^2+\frac{2 \sigma_3}{S_{min}\zeta_3}\left(\frac{{d^*_3} }{h^-_3}\right)^2+\frac{2\eta_3}{S_{min}\zeta_3}\left(\frac{S_{2} h_3}{h^-_3}\right)^2\\
\theta_4^*=& \frac{2 \upsilon_4}{ S_{min}\zeta_4}\left(\frac{\mu_4}{h^-_4}\right)^2+\frac{2 \sigma_4}{S_{min}\zeta_4}\left(\frac{{d^*_4} }{h^-_4}\right)^2+\frac{2\eta_4}{S_{min}\zeta_4}\left(\frac{S_{2} h_4}{h^-_4}\right)^2
\end{aligned}
\end{equation}
where $\upsilon_1, \ldots, \upsilon_4$, $\sigma_1, \ldots, \sigma_4$, $\delta_1$, $\eta_2$, $\eta_3$, and $\eta_4$ are unknown positive constants. Let us define the error of adaptive law:
\begin{equation}
\small
\begin{aligned}
\label{equaation: 25}
\tilde{\theta}_j=\hat{\theta}_j-\theta_j^*
\end{aligned}
\end{equation}
Now, we can rewrite \eqref{equaation: 25} using \eqref{equaation: 20} as follows:
\begin{equation}
\small
\begin{aligned}
\label{equaation: 26}
\dot{\tilde{\theta}}_j=-\beta_j \kappa_j \tilde{\theta}_j+\frac{1}{2} \zeta_j \beta_j \phi_j^2-\beta_j \kappa_j \theta_j^*
\end{aligned}
\end{equation}
\indent \textbf{Definition 2.} According to \cite{corless1993bounded} and \cite{heydari2024robust}, for $t \geq t_0$, the system tracking error ${x}_{ej}$ is uniformly exponentially bounded within a defined region $g\left(\tau\right)$ if:
\begin{equation}
\small
\begin{aligned}
\label{equaation: 27}
&\|x_{ej}\|=\|{x_j}-{x}_{jd}\| \leq \bar{c} e^{b (t-t_0)} \|{x}_{ej}(t_0)\| + \tilde{\mu}
\end{aligned}
\end{equation}
where $\bar{c}$, $\tilde{\mu}$ and  $b \in \mathbb{R}^+$ are positive constants. $x_{ej}(t_0)$ is any initial tracking error, and $g\left(\tau\right)$ can be defined as:
\begin{equation}
\small
\begin{aligned}
\label{equaation: 28}
&g\left(\tau\right):=\left\{{x}_{ej} \mid\|{x}_{ej} \| \leq {\tilde{\mu}}\right\}
\end{aligned}
\end{equation}
After making a derivative of \eqref{equation: 18}:
\begin{equation}
\small
\begin{aligned}
\label{equation: 29}
\dot{e}_{j} = \begin{cases} h_j e_{j+1}+h_j u_j+\bar{\Delta}_j +d_j & \hspace{0.2cm} j=1 \\
h_j S_1 {u_j}+\bar{\Delta}_j +d_j+h_j S_2 & \hspace{0.2cm} j=2,3,4 \end{cases}
\end{aligned}
\end{equation}
where:
\begin{equation}
\small
\begin{aligned}
\small
\label{equation: 30}
\bar{\Delta}_j = \begin{cases} \Delta_j-(1-h_j)\dot{x}_{jd} & \hspace{0.2cm} j=1 \\
\Delta_j-\dot{x}_{jd}-\dot{u}_1 & \hspace{0.2cm} j=2\\
\Delta_j-\dot{x}_{jd} & \hspace{0.2cm} j=3,4 \end{cases}
\end{aligned}
\end{equation}
\indent \textbf{Assumption 3}. Let $\mu_j$ and ${d_j}^* \in \mathbb{R}^+$ be unknown positive constants assigning the bound of uncertainties and disturbances, and let $r: \mathbb{R} \rightarrow \mathbb{R}^+$ be a continuously bounded function with strictly positive values if:
\begin{equation}
\small
\begin{aligned}
\label{equation: 31}
&\|\bar{\Delta}_j\| \hspace{0.1cm} \leq \mu_j r_j \hspace{0.1cm}, \hspace{0.2cm} \|{d_j}\| \hspace{0.1cm} \leq  {d_j}^*
\end{aligned}
\end{equation}
The control design is summarized in Algorithm \ref{alg:alg2}.\\
\indent \textbf{Theorem 1.} Consider the PMSM-driven EMLA system in \eqref{equation:5} with state constraints, as provided in \eqref{equaation: 17}, and control effort constraints in \eqref{equation: 8}. By employing the DRS-BLF control framework (Algorithm 1), all states of the EMLA system track the desired trajectories with uniformly exponential convergence.
\begin{algorithm}[H]
\scriptsize
\caption{Step-wise guidance of the DRS-BLF control}\label{alg:alg2}
\begin{algorithmic}
\STATE \textbf{Input:} $x_{1d}$, $x_{2d}$, current sensor information, and control constants.
\STATE \textbf{Output:} control signals $Sat(u_3)$ and $Sat(u_4)$.
\STATE
\STATE \hspace{0.0cm} \textbf{If} $j = 1$ \textbf{then}
\STATE \hspace{0.5cm} $x_{{e}_{j}}= x_{j} - x_{{{jd}}}$;
\STATE \hspace{0.5cm} $e_{j}= x_{j}$;
\STATE \hspace{0.5cm} $\phi_{j}=\frac{e_{j}}{o_{j}^2-e_{j}^2}$;
\STATE \hspace{0.5cm} $\hat{\theta}_{j}=-\beta_{j} \kappa_{j} \hat{\theta}_{j}+\frac{1}{2} \zeta_{j} \beta_{j} \phi_{j}^2$;
\STATE \hspace{0.5cm} $u_{j}=-\frac{1}{2}(\epsilon_{j} e_{j}+\zeta_{j} \hat{\theta}_{j} \phi_{j})$;
\STATE \hspace{0.0cm} \textbf{If}  $j = 2$ \textbf{then};
\STATE \hspace{0.5cm} $x_{{e}_{j}}= x_{j} - x_{{{jd}}}$;
\STATE \hspace{0.5cm} $e_{j}= x_{j}-u_{{j-1}}$;
\STATE \hspace{0.5cm} $\phi_{j}=\frac{e_{j}}{o_{j}^2-e_{j}^2}$;
\STATE \hspace{0.5cm} $\hat{\theta}_{j}=-\beta_{j} \kappa_{j} \hat{\theta}_{j_i}+\frac{1}{2} \zeta_{j} \beta_{j} \phi_{j}^2$;
\STATE \hspace{0.5cm} $u_{j}=-\frac{1}{2}(\epsilon_{j} e_{j}+\zeta_{j} \hat{\theta}_{j} \phi_{j})$;
\STATE \hspace{0.5cm} $Sat(u_{j})$;
\STATE \hspace{0.0cm} \textbf{Else} \textbf{then}
\STATE \hspace{0.5cm} $x_{{e}_{j}}= x_{j} - x_{{{jd}}}$;
\STATE \hspace{0.5cm} $e_{j}= x_{j}$;
\STATE \hspace{0.5cm} $\phi_{j}=\frac{e_{j}}{o_{j}^2-e_{j}^2}$;
\STATE \hspace{0.5cm} $\hat{\theta}_{j}=-\beta_{j} \kappa_{j} \hat{\theta}_{j}+\frac{1}{2} \zeta_{j} \beta_{j} \phi_{j}^2$;
\STATE \hspace{0.5cm} $u_{j}=-\frac{1}{2}(\epsilon_{j} e_{j}+\zeta_{j} \hat{\theta}_{j} \phi_{j})$;
\STATE \hspace{0.5cm} $Sat(u_{j})$;
\STATE \hspace{0.0cm} \textbf{end}
\end{algorithmic}
\label{alg2}
\end{algorithm}

\section{Uniformly Exponential Stability Analysis}
We introduce a BLF of the form:
\begin{equation}
\small
\begin{aligned}
\label{equaation: 32}
V_1=\frac{S_{min}}{2 h^-_1} \log \left(\frac{o_1^2}{Q_1}\right)+\frac{S_{min}}{2} \beta_1^{-1} \tilde{\theta}_1^2
\end{aligned}
\end{equation}
Then, by taking the derivative of \eqref{equaation: 32}:
\begin{equation}
\small
\begin{aligned}
\label{equaation: 33}
\dot{V}_1=&\frac{S_{min}h_1}{h^-_1} \phi_1 e_2 +\frac{S_{min}}{h^-_1} \phi_1 \bar{\Delta}_1+\frac{S_{min}h_1}{h^-_1} \phi_1 u_1\\
&+\frac{S_{min}}{h^-_1} \phi_1 d_1+S_{min}\beta_1^{-1} \tilde{\theta}_1\dot{\tilde{\theta}}_1
\end{aligned}
\end{equation}
Then, from \eqref{equation: 31}:
\begin{equation}
\small
\begin{aligned}
\label{equaation: 34}
\dot{V}_1 \leq& \frac{S_{min}h_1}{h^-_1} \phi_1 e_2+\frac{S_{min}}{h^-_1} \mu_1 r_1 \mid \phi_1 \mid+\frac{S_{min}h_1}{h^-_1} \phi_1 u_1 \\
&+\frac{S_{min}}{h^-_1} d^*_1 \mid \phi_1 \mid+S_{min}\beta_1^{-1} \tilde{\theta}_1\dot{\tilde{\theta}}_1 \\
\end{aligned}
\end{equation}
From Assumption 1, we know $h^+_1 \geq h_1$, and according to Young's inequality, by considering positive constants $\delta_1$, $\upsilon_1$, and $\sigma_1$, we can obtain:
\begin{equation}
\small
\begin{aligned}
\label{equaation: 35}
\dot{V}_1 \leq & S_{min}\delta_1\left(\frac{h^+_1}{h^-_1}\right)^2 \phi_1^2+\frac{1}{4} S_{min}\delta_1^{-1} e_2^2+\frac{1}{4} \upsilon_1^{-1} r_1^2\\
&+S_{min}^2\upsilon_1\left(\frac{\mu_1}{h^-_1}\right)^2  \phi_1^2 +\frac{S_{min}h_1}{h^-_1} \phi_1 u_1+\frac{1}{4} \sigma_1^{-1}\\
&  +S_{min}^2\sigma_1\left(\frac{{d^*_1} }{h^-_1}\right)^2 \phi_1^2+S_{min}\beta_1^{-1} \tilde{\theta}_1\dot{\tilde{\theta}}_1\\
\end{aligned}
\end{equation}
By considering the definition of $\theta^*_1$ from \eqref{equaation: 24}:
\begin{equation}
\small
\begin{aligned}
\label{equaation: 36}
\dot{V}_1 & \leq \frac{1}{2} S_{min} \zeta_1 \theta_1^* \phi_1^2 +\frac{1}{4} S_{min} \delta_1^{-1} e_2^2 +\frac{1}{4} \upsilon_1^{-1} r_1^2+\frac{1}{4} \sigma_1^{-1}\\
&+\frac{S_{min}h_1}{h^-_1} \phi_1 u_1+S_{min}\beta_1^{-1} \tilde{\theta}_1\dot{\tilde{\theta}}_1 \\
\end{aligned}
\end{equation}
Inserting $u_1$ and $\dot{\tilde{\theta}}_1$ from \eqref{equaation: 19} and \eqref{equaation: 26}:
\begin{equation}
\small
\begin{aligned}
\label{equaation: 37}
\dot{V}_1 \leq & \frac{1}{2} S_{min} \zeta_1 \theta_1^* \phi_1^2+\frac{1}{4} S_{min} \delta_1^{-1} e_2^2+\frac{1}{4} \upsilon_1^{-1} r_1^2 +\frac{1}{4} \sigma_1^{-1} \\
& -\frac{1}{2} \frac{S_{min} h_1}{h^-_1} \epsilon_1 \phi_1 e_1-\frac{1}{2} \frac{S_{min} h_1}{h^-_1} \zeta_1 \hat{\theta}_1 \phi_1^2 - S_{min} \kappa_1 \tilde{\theta}_1^2\\
&+\frac{1}{2} S_{min} \zeta_1 \tilde{\theta}_1 \phi_1^2 - S_{min} \kappa_1 \tilde{\theta}_1 \theta_1^* \\
\end{aligned}
\end{equation}
Because $h_1\geq h^-_1 > 0$, and $\tilde{\theta}_1=\hat{\theta}_1 - \theta_1^*$:\\
\begin{equation}
\small
\begin{aligned}
\label{equaation: 38}
\dot{V}_1 \leq & -\frac{1}{2} S_{min} \frac{h_1}{h^-_1} \epsilon_1 \phi_1 e_1  +\frac{1}{4} S_{min} \delta_1^{-1} e_2^2 -\frac{1}{2} S_{min} \kappa_1 \tilde{\theta}_1^2\\
& +\frac{1}{4} \upsilon_1^{-1} r_1^2-\frac{1}{2} S_{min} \kappa_1 (\hat{\theta}_1-\theta_1^*)^2 + S_{min} \kappa_1 {\theta_1^*}^2\\
&+\frac{1}{4} \sigma_1^{-1}
\end{aligned}
\end{equation}
and because $\hat{\theta}_1 > 0$:
\begin{equation}
\small
\begin{aligned}
\label{equaation: 39}
\dot{V}_1 \leq & -\frac{1}{2} S_{min} \frac{h_1}{h^-_1} \epsilon_1 \phi_1 e_1 +\frac{1}{4} \upsilon_1^{-1} r_1^2 +\frac{1}{4} S_{min} \delta_1^{-1} e_2^2 \\
&-\frac{1}{2} S_{min} \kappa_1 \tilde{\theta}_1^2 +\frac{1}{2} S_{min} \kappa_1 {\theta_1^*}^2 +\frac{1}{4} \sigma_1^{-1}
\end{aligned}
\end{equation}
According to Definition 1, \eqref{equaation: 39} can be rewritten as:
\begin{equation}
\small
\begin{aligned}
\label{equaation: 40}
\dot{V}_1 \leq & -\frac{1}{2} \frac{S_{min} h_1}{h^-_1} \epsilon_1 \log \left(\frac{\rho_1^2}{Q_1}\right) +\frac{1}{4} S_{min} \delta_1^{-1} e_2^2 \\
&  +\frac{1}{4} \upsilon_1^{-1} r_1^2-\frac{1}{2} S_{min} \kappa_1 \tilde{\theta}_1^2 +\frac{1}{2} S_{min} \kappa_1 {\theta_1^*}^2 +\frac{1}{4} \sigma_1^{-1}
\end{aligned}
\end{equation}
Regarding the definition of $V_1(\cdot)$ in \eqref{equaation: 32}:
\begin{equation}
\small
\begin{aligned}
\label{equaation: 41}
\dot{V}_1 \leq & -\psi_1 V_1 +\frac{1}{4} \upsilon_1^{-1} r_1^2 +\frac{1}{4} S_{min} \delta_1^{-1} e_2^2 +\frac{1}{2} S_{min} \kappa_1 {\theta_1^*}^2 \\
&+\frac{1}{4} \sigma_1^{-1}
\end{aligned}
\end{equation}
where:
\begin{equation}
\small
\begin{aligned}
\label{equaation: 42}
\psi_1 = \min \left[h_1 \epsilon_1, \kappa_1 \beta_1\right]
\end{aligned}
\end{equation}
Similarly, we suggest another BLF candidate, as follows:
\begin{equation}
\small
\begin{aligned}
\label{equaation: 43}
V_2=V_{1}+\frac{1}{2 h^-_2} \log \left(\frac{o_2^2}{Q_2}\right)+\frac{1}{2} S_{min} \beta_2^{-1} \tilde{\theta}_2^2
\end{aligned}
\end{equation}
After the derivative of \eqref{equaation: 43}, according to \eqref{equaation: 41} and \eqref{equation: 29}, we have:
\begin{equation}
\small
\begin{aligned}
\label{equaation: 44}
\dot{V}_2 \leq &-\psi_1 V_1 +\frac{1}{4} \upsilon_1^{-1} r_1^2 +\frac{1}{4} S_{min} \delta_1^{-1} e_2^2 +\frac{1}{2} S_{min} \kappa_1 {\theta_1^*}^2\\
&+\frac{1}{4} \sigma_1^{-1}+\frac{1}{h^-_2} \phi_2 \bar{\Delta}_2+\frac{S_1 h_2}{h^-_2} \phi_2 u_2+\frac{1}{h^-_2} \phi_2 d_2\\
&+\frac{S_2 h_2}{h^-_2} \phi_2+S_{min}\beta_2^{-1} \tilde{\theta}_2\dot{\tilde{\theta}}_2 \\
\end{aligned}
\end{equation}
Then, from \eqref{equation: 31}:
\begin{equation}
\small
\begin{aligned}
\label{equaation: 45}
\dot{V}_2 \leq&-\psi_1 V_1 +\frac{1}{4} \upsilon_1^{-1} r_1^2 +\frac{1}{4} S_{min} \delta_1^{-1} e_2^2 +\frac{1}{2} S_{min} \kappa_1 {\theta_1^*}^2\\
&+\frac{1}{4} \sigma_1^{-1}+ \frac{1}{h^-_2} \mu_2 r_2 \mid \phi_2 \mid+\frac{S_{1}h_2}{h^-_2} \phi_2 u_2   \\
&+\frac{S_2 h_2}{h^-_2} \mid \phi_2 \mid+\frac{1}{h^-_2} d^*_2 \mid \phi_2 \mid+S_{min}\beta_2^{-1} \tilde{\theta}_2\dot{\tilde{\theta}}_2 \\
\end{aligned}
\end{equation}
From \eqref{equation: 21}, considering $h^+_2 \geq h_2$ according to Assumption 1, Young's inequality, and positive constants $\upsilon_2$, $\sigma_2$, and $\eta_2$, we obtain:
\begin{equation}
\small
\begin{aligned}
\label{equaation: 46}
\dot{V}_2 \leq &-\psi_1 V_1 +\frac{1}{4} \upsilon_1^{-1} r_1^2 +\frac{1}{4} S_{min} \delta_1^{-1} Q_2^2 \phi_2^2 \\
&+\frac{1}{2} S_{min} \kappa_1 {\theta_1^*}^2+\frac{1}{4} \upsilon_2^{-1} r_2^2 +\upsilon_2\left(\frac{\mu_2}{h^-_2}\right)^2  \phi_2^2\\
&+\frac{1}{4} \sum_{i=1}^2 \sigma_i^{-1}+\frac{S_{1}h_2}{h^-_2} \phi_2 u_2+ \eta_2 \left(\frac{S_2 h_2}{h^-_2}\right)^2 \phi_2^2 +\frac{1}{4}  \eta_2^{-1}\\
&+\sigma_2\left(\frac{{d^*_2} }{h^-_2}\right)^2 \phi_2^2+S_{min}\beta_2^{-1} \tilde{\theta}_2\dot{\tilde{\theta}}_2
\end{aligned}
\end{equation}
By considering the definition of $\theta^*_2$ from \eqref{equaation: 24}:
\begin{equation}
\small
\begin{aligned}
\label{equaation: 47}
\dot{V}_2  \leq &-\psi_1 V_1 +\frac{1}{4} \sum_{i=1}^2 \upsilon_i^{-1} r_i^2 +\frac{1}{2} S_{min} \kappa_1 {\theta_1^*}^2\\
& +\frac{1}{2} S_{min} \zeta_2 \theta_2^* \phi_2^2+\frac{S_{1}h_2}{h^-_2} \phi_2 u_2 + \frac{1}{4} \eta_2^{-1}+\frac{1}{4} \sum_{i=1}^2 \sigma_i^{-1}\\
&+S_{min}\beta_2^{-1} \tilde{\theta}_2\dot{\tilde{\theta}}_2
\end{aligned}
\end{equation}
Inserting $u_2$ and $\dot{\tilde{\theta}}_2$ from \eqref{equaation: 19} and \eqref{equaation: 26}, knowing $S_{min}\leq S_1$:
\begin{equation}
\small
\begin{aligned}
\label{equaation: 48}
\dot{V}_2 \leq &-\psi_1 V_1 +\frac{1}{4} \sum_{i=1}^2 \upsilon_i^{-1} r_i^2  +\frac{1}{2} S_{min} \kappa_1 {\theta_1^*}^2+\frac{1}{2} S_{min} \zeta_2 \theta_2^* \phi_2^2\\
& -\frac{1}{2} S_{min} \zeta_2 \hat{\theta}_2 \phi_2^2+\frac{1}{4} \sum_{i=1}^2 \sigma_i^{-1} + \frac{1}{4} \eta_2^{-1}-\frac{1}{2} \frac{S_{1} h_2}{h^-_2} \epsilon_2 \phi_2 e_2 \\
& +\frac{1}{2} S_{min} \zeta_2 \tilde{\theta}_2 \phi_2^2- S_{min} \kappa_2 \tilde{\theta}_2^2- S_{min} \kappa_2 \tilde{\theta}_2 \theta_2^*
\end{aligned}
\end{equation}
Because $\tilde{\theta}_2=\hat{\theta}_2 - \theta_2^*$, and similar to \eqref{equaation: 38}:\\
\begin{equation}
\small
\begin{aligned}
\label{equaation: 49}
\dot{V}_2 \leq &-\psi_1 V_1 +\frac{1}{4} \sum_{i=1}^2 \upsilon_i^{-1} r_i^2  +\frac{1}{2} S_{min} \kappa_1 {\theta_1^*}^2+\frac{1}{4} \sum_{i=1}^2 \sigma_i^{-1}\\
&  + \frac{1}{4} \eta_2^{-1}-\frac{1}{2} \frac{S_{1} h_2}{h^-_2} \epsilon_2 \phi_2 e_2- S_{min} \kappa_2 \tilde{\theta}_2^2- S_{min} \kappa_2 \tilde{\theta}_2 \theta_2^*
\end{aligned}
\end{equation}
According to Definition 1, and similar to \eqref{equaation: 40}:
\begin{equation}
\small
\begin{aligned}
\label{equaation: 50}
\dot{V}_2 \leq &-\psi_1 V_1 +\frac{1}{4} \sum_{i=1}^2 \upsilon_i^{-1} r_i^2  +\frac{1}{2}  \sum_{i=1}^2 \kappa_i {\theta_i^*}^2+\frac{1}{4} \sum_{i=1}^2 \sigma_i^{-1}\\
&+\frac{1}{4}\eta_2^{-1} -\frac{1}{2} \frac{S_1 h_2}{h^-_2} \epsilon_2 \log \left(\frac{\rho_2^2}{Q_2}\right)-\frac{1}{2} S_{min} \kappa_2 \tilde{\theta}_2^2
\end{aligned}
\end{equation}
By considering the definition of $V_2(\cdot)$ in \eqref{equaation: 43}, and $S_1>0$:
\begin{equation}
\small
\begin{aligned}
\label{equaation: 51}
\dot{V}_2 \leq & -\psi_2 V_2 +\frac{1}{4} \sum_{i=1}^2 \upsilon_i^{-1} r_i^2 +\frac{1}{2} \sum_{i=1}^2 \kappa_i {\theta_i^*}^2 +\frac{1}{4} \sum_{i=1}^2 \sigma_i^{-1}\\
&+\frac{1}{4}\eta_2^{-1}
\end{aligned}
\end{equation}
where:
\begin{equation}
\small
\begin{aligned}
\label{equaation: 52}
\psi_2 = \min \left[\psi_1, S_1 h_2 \epsilon_2, \kappa_2 \beta_2\right]
\end{aligned}
\end{equation}
Similarly, we can assume two other BLF candidates, as:
\begin{equation}
\small
\begin{aligned}
\label{equaation: 53}
V_3&=V_{2}+\frac{1}{2 h^-_3} \log \left(\frac{o_3^2}{Q_3}\right)+\frac{1}{2} S_{min} \beta_3^{-1} \tilde{\theta}_3^2\\
V_4&=V_{3}+\frac{1}{2 h^-_4} \log \left(\frac{o_4^2}{Q_4}\right)+\frac{1}{2} S_{min} \beta_4^{-1} \tilde{\theta}_4^2
\end{aligned}
\end{equation}
As with the first two Lyapunov functions, we can achieve:
\begin{equation}
\small
\begin{aligned}
\label{equaation: 54}
\dot{V}_3 \leq & -\psi_3 V_3 +\frac{1}{4} \sum_{i=1}^3 \upsilon_i^{-1} r_i^2 +\frac{1}{2} \sum_{i=1}^3 \kappa_i {\theta_i^*}^2 +\frac{1}{4} \sum_{i=1}^3 \sigma_i^{-1} \\
&+\frac{1}{4} \sum_{i=2}^3 \eta_i^{-1}
\end{aligned}
\end{equation}
where:
\begin{equation}
\small
\begin{aligned}
\label{equaation: 55}
\psi_3 = \min \left[\psi_1, \psi_2, S_1 h_3 \epsilon_3, \kappa_3 \beta_3\right]
\end{aligned}
\end{equation}
and similarly:
\begin{equation}
\small
\begin{aligned}
\label{equaation: 56}
\dot{V}_4 \leq & -\psi_4 V_4 +\frac{1}{4} \sum_{i=1}^4 \upsilon_i^{-1} r_i^2 +\frac{1}{2} \sum_{i=1}^4 \kappa_i {\theta_i^*}^2 +\frac{1}{4} \sum_{i=1}^4 \sigma_i^{-1} \\
&+\frac{1}{4} \sum_{i=2}^4 \eta_i^{-1}
\end{aligned}
\end{equation}
where:
\begin{equation}
\small
\begin{aligned}
\label{equaation: 57}
\psi_4 = \min \left[\psi_1, \psi_2, \psi_3, S_1 h_4 \epsilon_4, \kappa_4 \beta_4\right]
\end{aligned}
\end{equation}
Next, considering $V = V_4$, we rewrite \eqref{equaation: 56} as follows:
\begin{equation}
\small
\begin{aligned}
\label{equaation: 59}
\dot{V}\leq & -\psi_4 V_4 +\frac{1}{4} \sum_{i=1}^4 \upsilon_i^{-1} r_i^2 +\mu^*
\end{aligned}
\end{equation}
where:
\begin{equation}
\small
\begin{aligned}
\label{equaation: 60}
\mu^*= \frac{1}{2} \sum_{i=1}^4 \kappa_i {\theta_i^*}^2 +\frac{1}{4} \sum_{i=1}^4 \sigma_i^{-1}+\frac{1}{4} \sum_{i=2}^4 \eta_i^{-1}
\end{aligned}
\end{equation}
From \eqref{equaation: 59}, we can reach:
\begin{equation}
\small
\begin{aligned}
\label{equaation: 61}
V \leq & V\left(t_0\right) e^{-\psi_4\left(t-t_0\right)} +\frac{1}{4} \sum_{i=1}^4 \nu_i^{-1} \int_{t_0}^t e^{-\psi_4(t-T)} r_i^2 \mathrm{d} T \\
& +\mu^* \int_{t_0}^t e^{-\psi_4(t-T)} \mathrm{d} T
\end{aligned}
\end{equation}
According to \cite{heydari2024robust}, following Definition 2, we can infer that the studied PSMS-powered EMLA states remain uniformly bounded within a ball region, the radius of which is subject to external disturbances, as well as system uncertainties.
Presently, we examine the constraints on the system states $x_j(t)$ within the set $\chi_j$, ensuring they are preserved. Considering the definition of the state transformation in \eqref{equation: 18} and the constraints provided in \eqref{equaation: 17}, we can accomplish this for the EMLA system, as follows:
\begin{equation}
\small
\begin{aligned}
\label{equaation: 66}
 \left|x_j\right| \leq\left|e_j\right|+\left|x_{jd}\right|<\rho_j+\lambda_j=\chi_{j}-\lambda_j+\lambda_j=\chi_{j}
 \end{aligned}
\end{equation}
The virtual control $u_1$ is influenced by all bounded signals (namely, $x_1$ and $\hat{\theta}_1$), indicating the uniform boundedness of the virtual control signal $u_1$. 
\eqref{equaation: 66} asserts that the system states $x_j$ consistently stays within the set $\chi_j$. 
\section{Optimization Of Drs-blf Control Parameters}
The Jaya algorithm can optimize gains through simple principles, moving toward the best solution while avoiding the worst and requiring minimal parameter tuning \cite{rao2019JAYA}. The operational process of the Jaya algorithm for DRS-BLF control gains is outlined in Algorithm \ref{alg:alg3}.
\begin{algorithm}[H]
\scriptsize
\caption{Jaya optimization procedure.}\label{alg:alg3}
\begin{algorithmic}
\STATE 
\STATE {\textsc{\textbf{Initialize random positive candidates:}}}
\STATE \hspace{0.5cm}population size: $n_c \in \mathbb{R}$;
\STATE \hspace{0.5cm}vector of the candidate: $c_\ell \in \mathbb{R^\textit{N}}$;
\STATE \hspace{0.5cm}number of control gains: $N\in \mathbb{R}$;
\STATE \hspace{0.5cm}objective function: $f_x \in \mathbb{R}$;
\STATE 
\STATE \textsc{\textbf{find}} the best solution set such that $\min[f_x(c_\ell)]=f_x(c_{best})$;
\STATE \textsc{\textbf{find}} the worst solution set such that $\max[f_x(c_\ell)]=f_x(c_{worst})$;
\STATE
\STATE {\textsc{\textbf{For}}} $\ell=1:n_c$
\STATE \hspace{0.5cm}Modify all candidates by:
\STATE \hspace{1cm}$c_{\ell}(new) = c_\ell + r_{\ell1}(c_{best} - c_\ell) - r_{\ell2}(c_{worst} - c_\ell)$;
\STATE \hspace{1cm}if $c_{\ell}(new) \leq 0$:
\STATE \hspace{1.5cm}repeat {\scriptsize$c_{\ell}(new) = c_\ell + r_{\ell1}(c_{best} - c_\ell) - r_{\ell2}(c_{worst} - c_\ell)$};
\STATE \hspace{1cm}if $f_x(c_{\ell}(new))<f_x(c_\ell)$:
\STATE \hspace{1.5cm}replace $c_\ell$ with $c_\ell(new)$;
\STATE \hspace{1cm}Else:
\STATE \hspace{1.5cm}No change in the candidate;
\STATE \hspace{1.0cm}Replace new best ($c_{best}$) and worst candidates ($c_{worst}$);
\STATE \hspace{0.0cm}\textsc{\textbf{end}}
\STATE
\STATE \hspace{0.0cm}\textsc{\textbf{Display}} the optimal gains ($c_{best}$);
\end{algorithmic}
\label{alg1}
\end{algorithm}
As outlined, the primary goal is to minimize an objective function. In this paper, for the DRS-BLF control employed for the EMLA, the objective function ($f_{x}$) is selected based on position and velocity tracking errors, as follows:
\begin{equation}
\small
\begin{aligned}
\label{equaation: 67}
f_{x}=\sqrt{\|x_{e_{1}}\|^2+\|x_{e_{2}}\|^2}
 \end{aligned}
\end{equation}
\begin{figure*} [t]
    \centering
    \includegraphics[width=0.95\textwidth, height=9.5cm]{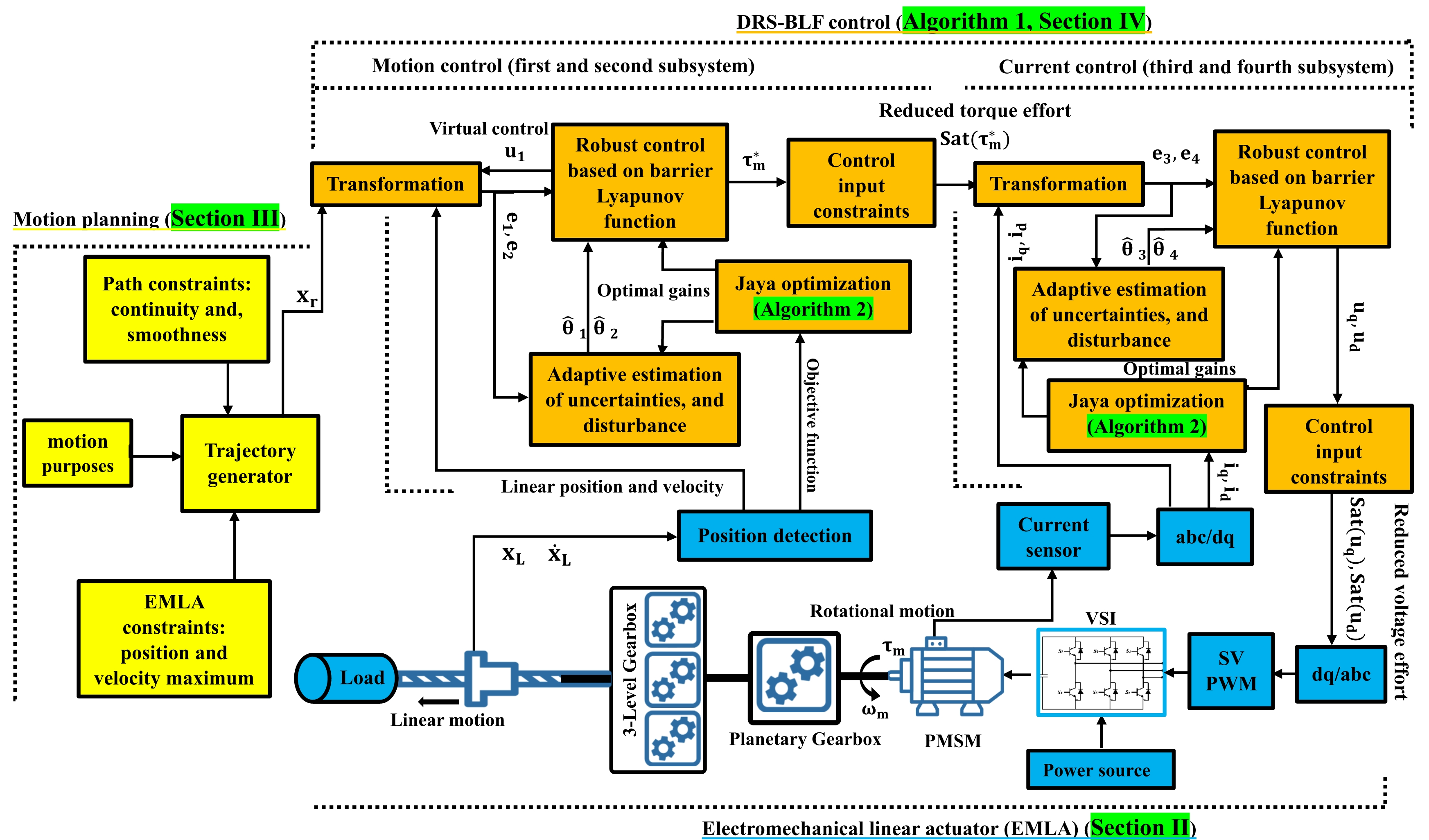}
    \caption{Interconnections of the EMLA mechanism (blue boxes), implemented by motion planning (yellow boxes) and DRS-BLF control (orange boxes).}
    \label{fig1111}
\end{figure*}
The updating of candidate $c_\ell$ involves the utilization of two random vectors, $r_{\ell1}$ and $r_{\ell2}$, with N elements within the range of $[0,1]$. The expressions $r_{1}(c_{best} - c_\ell)$ and $- r_{2}(c_{worst} - c_\ell)$ indicate the algorithm's movement toward the best solution while avoiding the worst. Given the DRS-BLF control gains, denoted as $\beta_j$, $\kappa_j$, $\zeta_j$, and $\epsilon_j$ for the four subsystems, the full number of control parameters is $16$.

\begin{table}[h!]
  \caption{Specifications of the studied PMSM}
  \vspace{-0.5cm}
  \begin{center}
  \begin{tabular}{|c|c|}
    \hline
    \textbf{PMSM Specification} & \textbf{Value} \\
    \hline
    IP Rating & IP 66 \\
    \hline
    Insulation & Class F \\
    \hline
    Interface & BiSS/EnDat 4.8-10 V \\
    \hline
    Continuous Torque (Mc) & 77 Nm @ 48.2 A \\
    \hline
    Nominal Torque (Mn) & 37 Nm @ 23.1 A \\
    \hline
    Safe Maximum Torque (Mcs) & 98 Nm @ 48.2 A \\
    \hline
    The Back EMF Constant (Ke) & 81.7 Vpk/krpm \\
    \hline
    Torque Constant (Kt) & 16 Nm/A \\
    \hline
    Cs @ 140C & 6.49 A P: 11.6 kW \\
    \hline
    Nmax & 3000 rpm / 3877 rpm \\
    \hline
    Drive VPWM & 380/480 VAC \\
    \hline
    Brake & N/A \\
    \hline
  \end{tabular}
  \label{pmsm}
  \end{center}
\end{table}

\begin{figure} [h!]
    \centering
    \scalebox{1}{\includegraphics[trim={0cm 0.0cm 0.0cm 0cm},clip,width=\columnwidth]{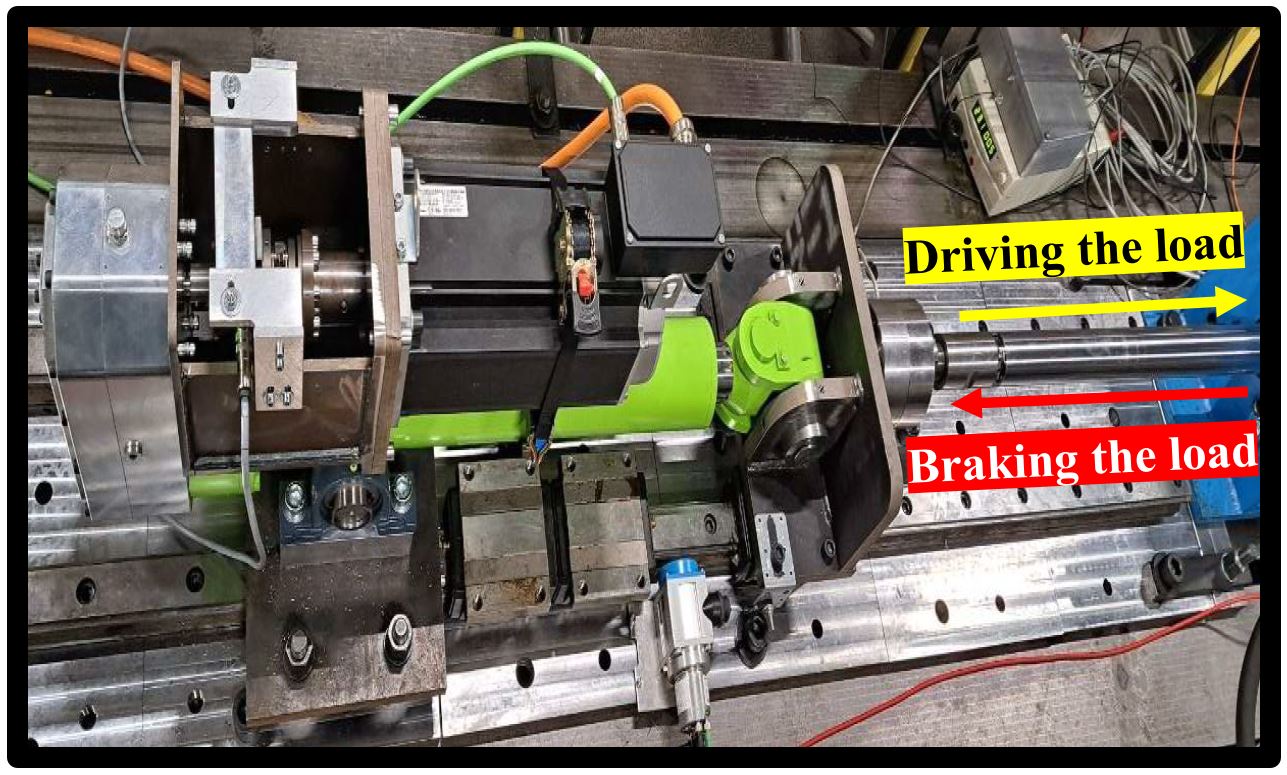}}
    \caption{The studied PMSM-driven EMLA under a pushing load generated by an external linear hydraulic actuator.}
    \label{fig11}
\end{figure}

\section{Experimental Results}
\begin{figure}[h!]
    \hspace*{-0.0cm} % Adjust the value as needed
    \centering
    \scalebox{1}{\includegraphics[trim={0cm 0.0cm 0.0cm 0cm},clip,width=\columnwidth]{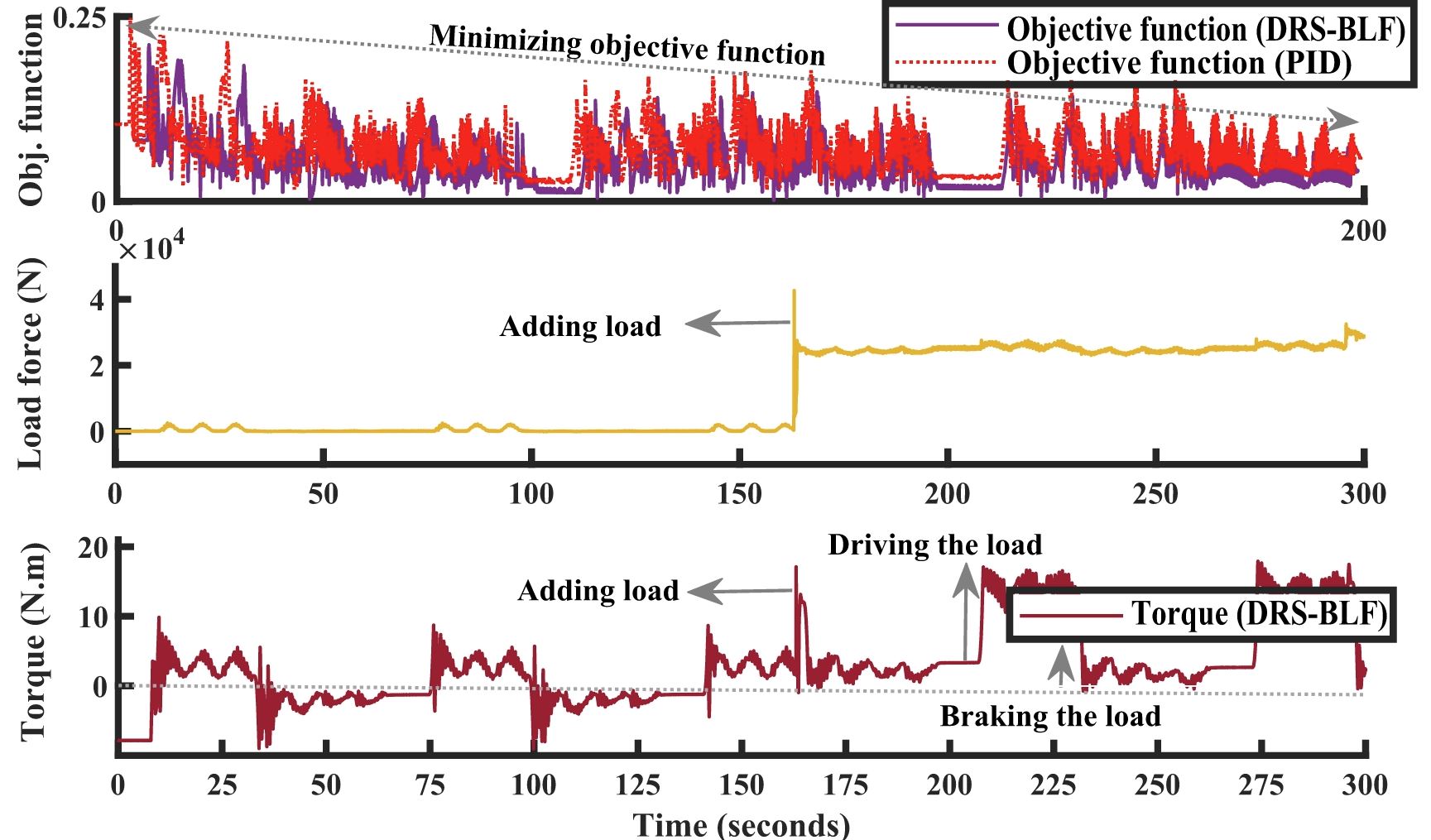}}
    \caption{(a) Minimizing objective function, (b) load force generated by a hydraulic actuator, (c) torque generated by the EMLA.}
    \label{fig6}
\end{figure}

\begin{figure}[h!]
    \hspace*{-0.0cm} % Adjust the value as needed
    \centering
    \scalebox{1}{\includegraphics[trim={0cm 0.0cm 0.0cm 0cm},clip,width=\columnwidth]{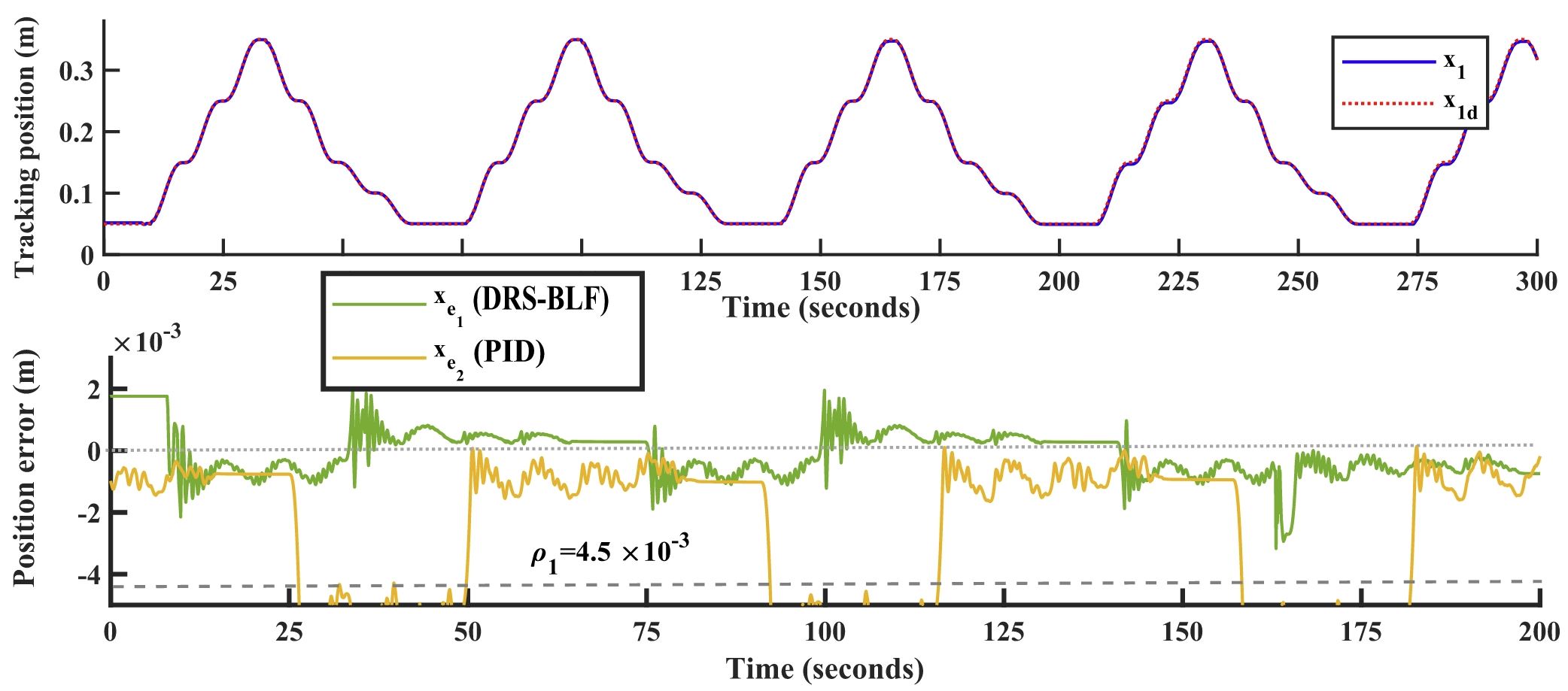}}
    \caption{Position-tracking performance of the studied EMLA, determined by employing the DRS-BLF and PID control.}
    \label{fig7}
\end{figure}

The block diagram shown in Fig. \ref{fig1111} illustrates the interconnection among different sections of the PMSM-powered EMLA system and DRS-BLF control sections. The DRS-BLF controller is implemented within the Unidrive M700 controller (model number: M700-064 00350 A), tasked with executing the control logic to manage the servo motor's operations effectively. The Unidrive M700 controller is connected to an inverter that drives the servo motor. Communication and control commands between the controller and inverter, as well as other system components, are facilitated through an EtherCAT network, allowing for real-time control and monitoring and ensuring the motion control is precise and adheres to the safety and performance criteria set by the control framework.
The specifications of the used PMSM are indicated in Table \ref{pmsm}, and an image of the studied EMLA is depicted in Fig. \ref{fig11}.
The sample time of the DRS-BLF control system was set to 1,000 Hz.
The torque generated by the DRS-BLF was directly obtained from the controller through a 16-bit message, and the studied EMLA was equipped with an external linear sensor (SICK BTF08, resolution: 18 bits, range: 1.5 m) to measure the linear position of the EMLA. 
By employing quantic polynomials from Chapter 13 of \cite{jazar2010theory}, a jerk-bounded path was generated as the reference trajectory, selected because it is a comprehensive reference trajectory encompassing forward, stationary, and backward movements.
Initially, the DRS-BLF control was employed for the PMSM-driven EMLA to operate without any load for approximately 165 s to track the jerk-bounded reference trajectory. After approximately 165 s, we applied a rather constant load of 25 kN, generated by an external hydraulic actuator, in the direction of the EMLA movement while it tracked the reference trajectory. This load force was directed inward, resulting in a driving force from the EMLA when moving forward and a braking force in the reverse position (see  Fig. \ref{fig11}). 
Fig. \ref{fig6} indicates that by employing the Jaya algorithm, the objective function, as defined in \eqref{equaation: 67}, begins minimizing 2 s after beginning the task, and optimal gains are obtained as follows:\\
\begin{equation}
\small
\begin{aligned}
\label{equaation: 100}
&\beta_1=11.2, \beta_2=19.8, \beta_3=4.75, \beta_4=7.1\\
&\kappa_1=98, \kappa_2=76, \kappa_3=24, \kappa_4=48\\
&\zeta_1=0.002, \zeta_2=0.001, \zeta_3=0.0001, \zeta_4=0.0021\\
&\epsilon_1=0.005,\epsilon_2=0.008,\epsilon_3=0.001,\epsilon_4=0.003
 \end{aligned}
\end{equation}
\\
In this regard, the Jaya parameters for tuning the DRS-BLF control were defined for the population size $n_c=15$ and the number of gains $N=16$. In addition, the second image in Fig. \ref{fig6} indicates the value of the load force generated by the hydraulic actuator after 165 s.
The third image in the figure illustrates the torque exerted by the EMLA for driving and braking the imposed linear load. While adhering to the nominal torque value specified in the PMSM's specifications, it becomes evident that upon adding a load, residual positive values of approximately $5$ N.m persisted to sustain the load's position, which exerted an inward force on the EMLA (in braking mode).
Fig. \ref{fig7} depicts the effectiveness of the DRS-BLF control in position tracking. The figure demonstrates, unlike proportional-integral-derivative (PID) control, the effectiveness of the safety concept within the DRS-BLF control for the studied EMLA, particularly with a small $\rho_1$ value.

\begin{figure}[h!]
    \hspace*{-0.0cm} % Adjust the value as needed
    \centering
    \scalebox{1}{\includegraphics[trim={0cm 0.0cm 0.0cm 0cm},clip,width=\columnwidth]{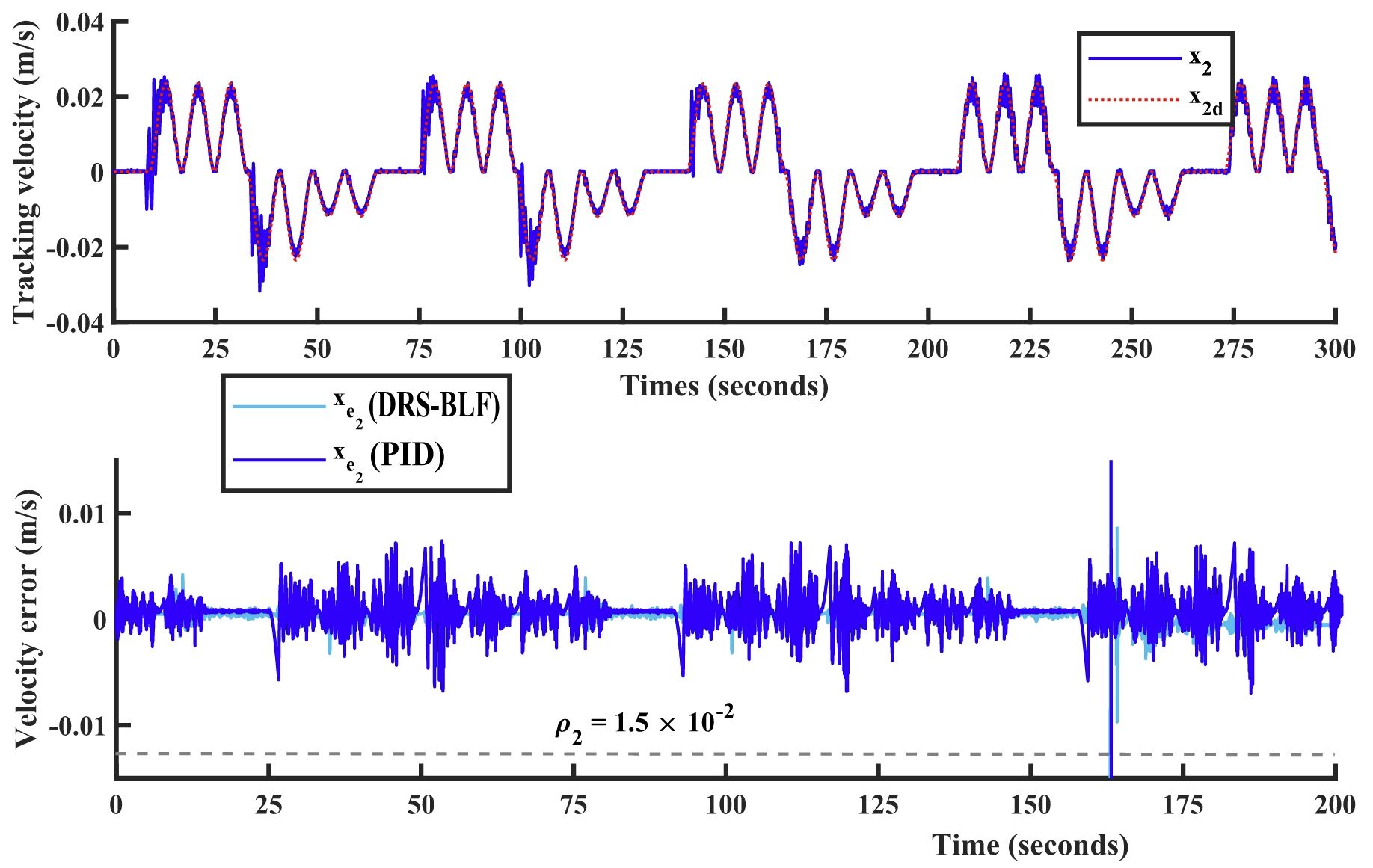}}
    \caption{Velocity-tracking performance of the studied EMLA, determined by employing the DRS-BLF and PID control.}
    \label{fig8}
\end{figure}

\begin{table}[htbp]
  \caption{Performance of Jaya-optimized DRL-BLF and PID control.}
  \vspace{-0.5cm}
  \begin{center}
  \begin{tabular}{|c|c|c|}
    \hline
    \textbf{Convergence Criteria} & \textbf{DRS-BLF Control} & \textbf{PID Control} \\
    \hline
    {Position error (m)} & 0.0035 & 0.0047 \\
    \hline
    {Velocity error (m/s)} & 0.0072 & 0.0086 \\
    \hline
    {Torque effort (N.m)} & 14.2 & 15.7 \\
    \hline
    {Convergence speed (s)} & 1.98 & 2.32 \\
    \hline
  \end{tabular}
  \label{tab:performance_comparison}
  \end{center}
\end{table}

The accuracy of the linear velocity in tracking the reference trajectory is demonstrated in Fig. 
\ref{fig8}, with a small $\rho_2$ value. Interestingly, the loaded EMLA exhibited a better performance compared to the unloaded performance in velocity tracking, likely owing to the robustness of the proposed control method or to the load capacity of the PMSM.  
Except for considering safety criteria and utilizing the auto-tuning Jaya method, experimental outcomes are summarized in Table \ref{tab:performance_comparison}.
This table demonstrates an improvement in control performance by employing the DRL-BLF control compared to the results obtained by PID control under the same conditions for the studied PMSM-driven EMLA. Specifically, there is an approximately $15\%$ improvement in the accuracy of position and velocity tracking, a $9\%$ decrease in torque efforts, and a $10\%$ faster convergence into the reference trajectories.
\section{Conclusion}
In this study, we presented an innovative DRS-BLF control framework for a PMSM-powered EMLA mechanism aimed at tracking reference trajectories while managing user-specified safety constraints. In addition, this approach employed a swarm intelligence technique to optimize DRS-BLF control parameters for minimizing tracking errors. By employing the DRS-BLF control framework, we guaranteed the robustness of the studied EMLA while also proving uniformly exponential stability.
The effectiveness of the DRS-BLF control was validated through experimental results compared with the PID control results, demonstrating an improvement in control tracking performance.

\bibliographystyle{IEEEtran}
\bibliography{ref}

\end{document}